\documentstyle[12pt]{article}
\begin{document}

\begin{titlepage}
\setcounter{page}{1}
\title{\bf Non-separability without Non-separability in Nonlinear Quantum 
Mechanics}
\author{Waldemar Puszkarz\thanks{Electronic address: puszkarz@cosm.sc.edu}
\\
\small{\it  Department of Physics and Astronomy,}
\\
\small{\it University of South Carolina,}
\\
\small{\it Columbia, SC 29208}}
\date{\small (May 15, 1999)}
\maketitle
\begin{abstract}
%\double
We show an example of benign non-separability in an apparently separable 
system consisting of $n$ free non-correlated quantum particles, solitonic 
solutions to the nonlinear phase modification of the Schr\"{o}dinger equation 
proposed recently. The non-separability manifests itself in the wave function 
of a single particle being influenced by the very presence of other particles. 
In the simplest case of identical particles, it is the number of particles that 
affects the wave function of each particle and, in particular, the width of 
its Gaussian probability density. As a result, this width, a local property, 
is directly linked to the mass of the entire Universe in a very Machian manner. 
In the realistic limit of large $n$, if the width in question is to be 
microscopic, the coupling constant must be very small resulting in an 
``almost linear'' theory. This provides a model explanation of why the 
linearity of quantum mechanics can be accepted with such a high degree of 
certainty even if the more fundamental underlying theory could be nonlinear. 
We also demonstrate that when such non-correlated solitons are coupled to 
harmonic oscilators they  lead to a faster-than-light nonlocal telegraph 
since changing the frequency of one oscillator affects instantaneously the 
probability density of particles associated with other oscillators. This 
effect can be alleviated by fine-tuning the parameters of the solution, which 
results in nonlocal correlations that affect only the phase. Exclusion rules 
of a novel kind that we term supersuperselection rules also emerge from 
these solutions. They are similar to the mass and the univalence 
superselection rules in linear quantum mechanics, but derive from different 
assumptions. The effects in question and the exclusion rules do not appear 
if a weakly separable extension to $n$-particles is employed.

\vskip 0.5cm
\noindent
%PACS numbers: 11.80.Et, 12.10.Gq, 04.90.$+$e
\end{abstract}

\end{titlepage}

\section{Introduction}

%%\double

Recently, we have presented the nonlinear phase modification of the
Schr\"{o}dinger equation called the simplest minimal phase extension (SMPE)
of the Schr\"{o}dinger equation \cite{Pusz1} together with some
one-dimensional solutions to it \cite{Pusz2}. The most interesting of them
is a free solitonic Gaussian solution describing a quantum particle. By
assuming that the self-energy term constitutes the rest mass-energy of the
soliton, we obtained a model particle whose physical size defined as the
width of its Gaussian probability density is equal to its Compton
wavelength. In this approach, that we call subrelativistic, the coupling
constant is inevitably a particle-dependent quantity, an intrinsic property
of the particle not unlike its mass. In fact, it is a function of the mass.
In a more general approach in which no such use is made of the self-energy
term, the size of the particle is determined both by its mass and by the
coupling constant, the latter being entirely independent of the former.

The main purpose of this paper is to present a free $n$-particle solution to
the modification in question describing uncorrelated solitons and to
demonstrate some unusual property of it that one could call the
non-separability without non-separability. As we will see, the $n$-particle
equation can be separated into $n$ one-particle equations, but some residual
non-separability that manifests itself in the dependence of individual
particle's wave function on the presence and properties of other particles
in the system remains. In addition, we will present a solution describing $n$
uncorrelated particles coupled to linear harmonic oscillators and discuss
the non-separability that occurs in this case which, in general, is less
benign than the non-separability of free particles. It is the analysis of
this case that demonstrates that a different $n$-particle extension from the
one we concentrate on in this paper is necessary to avoid pathological
situations in which the nonlocality is manifestly compromised. We suggest
such an extension for factorizable wave functions.

In what follows, $R$ and $S$ denote the amplitude and the phase of the wave
function\footnote{%
We follow here the convention of \cite{Pusz1} that treats the phase as the
angle. In a more common convention $S$ has the dimensions of action and $%
\Psi =Rexp(iS/\hbar )$.} $\Psi =Rexp(iS)$, $C$ is the only constant of the
modification which does not appear in linear quantum mechanics.

For $n$ particles of mass $m_{i}$, the equations of motion for the
modification read

\begin{equation}
\hbar \frac{\partial R^{2}}{\partial t}+\sum_{i=1}^{n}\frac{\hbar ^{2}}{m_{i}%
}\vec{\nabla}_{i}\cdot \left( R^{2}\vec{\nabla}_{i}S\right)
-2C\sum_{i=1}^{n}\Delta _{i}\left( R^{2}\sum_{i=1}^{n}\Delta _{i}S\right) =0,
\label{1}
\end{equation}
\begin{equation}
\sum_{i=1}^{n}\frac{\hbar ^{2}}{m_{i}}\Delta _{i}R-2R\hbar \frac{\partial S}{%
\partial t}-2RV-\sum_{i=1}^{n}\frac{\hbar ^{2}}{m_{i}}R\left( \vec{\nabla}%
_{i}S\right) ^{2}-2CR\left( \sum_{i=1}^{n}\Delta _{i}S\right) ^{2}=0,
\label{2}
\end{equation}
and the energy functional is 
\begin{equation}
E=\int \,d^{3}x\,\left\{ \sum_{i=1}^{n}\frac{\hbar ^{2}}{2m_{i}}\left[
\left( \vec{\nabla}_{i}R\right) ^{2}+R^{2}\left( \vec{\nabla}_{i}S\right)
^{2}\right] +CR^{2}\left( \sum_{i=1}^{n}\Delta _{i}S\right)
^{2}+VR^{2}\right\}.  \label{3}
\end{equation}
These can be derived from the Lagrangian density proposed in \cite{Pusz1}
and generalized to $n$ particles as follows 
\begin{equation}
-L_{n}(R,S)=\hbar R^{2}\frac{\partial S}{\partial t}+\sum_{i=1}^{n}\frac{%
\hbar ^{2}}{2m_{i}}\left[ \left( \vec{\nabla}_{i}R\right) ^{2}+R^{2}\left( 
\vec{\nabla}_{i}S\right) ^{2}\right] +CR^{2}\left( \sum_{i=1}^{n}\Delta
_{i}S\right) ^{2}+R^{2}V.  \label{4}
\end{equation}
The energy functional is a conserved quantity for the potentials $V$ that do
not depend explicitly on time and coincides with the quantum-mechanical
energy defined as the expectation value of the Hamiltonian for this
modification \cite{Pusz1, Pusz3}.

\section{ N-Particle Solution to the SMPE}

As demonstrated in \cite{Pusz2}, the modification under discussion possesses
a free solitonic solution. Its amplitude is that of a Gaussian, 
\begin{equation}
R(x,t)=N\exp \left[ -\frac{(x-vt)^{2}}{s^{2}}\right] ,  \label{5}
\end{equation}
where $v$ is the speed of the particle and $s$ is the half-width of the
Gaussian to be determined through the coupling constant $C$ and other
fundamental constants of the modification. The normalization constant $%
N=\left( 2/\pi s^{2}\right) ^{1/{4}}$. The phase of the soliton has the form 
\begin{equation}
S(x,t)=a(x-vt)^{2}+bvx+c(t),  \label{6}
\end{equation}
where $a$ and $b$ are certain constants and $c(t)$ is a function of time,
all of which, similarly as $v$ and $s$, are to be found from the equations
of motion. These equations are satisfied provided 
\begin{equation}
b=m/\hbar ,s^{2}=-8mC/\hbar ^{2},s^{4}a^{2}=1,  \label{7}
\end{equation}
and 
\begin{equation}
2\hbar s^{4}m\frac{\partial c(t)}{\partial t}+2\hbar ^{2}s^{2}+\hbar
^{2}s^{4}b^{2}v^{2}+8Ca^{2}s^{4}m=0.  \label{8}
\end{equation}
We see that the coupling constant $C$ has to be negative, $C=-|C|$. The
energy of the particle turns out to be 
\begin{equation}
E=E_{st}+\frac{mv^{2}}{2},  \label{9}
\end{equation}
where 
\begin{equation}
E_{st}=\frac{\hbar ^{2}}{2ms^{2}}=\frac{\hbar ^{4}}{16m^{2}\left| C\right| }
\label{10}
\end{equation}
is the stationary part of it.

Before we present the solitonic solution for $n$ uncorrelated solitons that
generalizes the above one, let us first mention the existence of a two
particle solution that differs from this $n$-soliton solution when the
latter is specified to the two-particle case. This is so because the
solution in question is not factorizable for $x$ and $y$. If we assume that
the particles have the same mass $m$, their ``entangled'' configuration is
given by the amplitude and the phase of wave function as follows 
\begin{equation}
R(x,y,t)=N_{2}\exp \left[ -\frac{(x-y-vt)^{2}}{s^{2}}\right] \exp \left[ -%
\frac{(x+y\mp vt)^{2}}{s^{2}}\right] ,  \label{11}
\end{equation}
\begin{equation}
S(x,y,t)=a\left[ (x-y-vt)^{2}+(x+y\mp vt)^{2}\right] +v\left[
b_{-}(x-y)+b_{+}(x+y)\right] +c(t),  \label{12}
\end{equation}
where $s^{2}=-32Cm/\hbar ^{2}$, $a^{2}=1/s^{4}$ and $b_{-}=\mp
b_{+}=m/2\hbar $ in accord with the upper/lower signs $\mp $ in these
formulas. We note that the term $-\left[ (x-y-vt)^{2}+(x+y\mp vt)^{2}\right]
/s^{2}$, that appears in the amplitude, cannot be obtained as a linear
combination of $-(x-vt)^{2}/s^{2}$ and $-(y\mp vt)^{2}/s^{2}$ which we will
use to build the amplitude for two particles in the general case of
uncorrelated solitons.\footnote{%
On the other hand, the amplitudes containing only $-(x-y\pm vt)^{2}/s^{2}$
or $-(x+y\pm vt)^{2}/s^{2}$ would give rise to wave functions that are not
normalizable.} This confirms that the solution (11-12) cannot be factorized
for $x$ and $y$. The energy of this configuration is $E=mv^{2}/2+\hbar
^{4}/16m^{2}\left| C\right| $ with the kinetic energy being, surprisingly
enough, just as for one particle. We did not succeed in finding similar
solutions in the presence of potentials. The most obvious choice, the
potential of harmonic oscillator does not support solutions of this type. It
does support though the solutions that we will discuss now.

To find the already mentioned $n$-soliton solution, we will proceed in the
same manner as for the one particle solution. To begin with, we assume that
the solitons are one-dimensional, but we will remove this condition later
on. The amplitude for the system of $n$ such solitons is chosen by analogy
to (5) as 
\begin{equation}
R_{n}(x_{1},x_{2},...,x_{n},t)=N_{n}\prod_{i=1}^{n}\exp \left[ -\frac{%
(x_{i}-v_{i}t)^{2}}{s_{i}^{2}}\right] ,  \label{13}
\end{equation}
and its phase is taken in the form 
\begin{equation}
S_{n}(x_{1},x_{2},...,x_{n},t)=\sum_{i=1}^{n}\left[
a_{i}(x_{i}-v_{i}t)^{2}+b_{i}v_{i}x_{i}\right] +c(t).  \label{14}
\end{equation}
The energy equation (2) is separable if and only if 
\begin{equation}
s_{i}^{4}=\frac{1}{a_{i}^{2}},  \label{15}
\end{equation}
and 
\begin{equation}
b_{i}=\frac{m_{i}}{\hbar },  \label{16}
\end{equation}
whereas the continuity equation (1) requires that 
\begin{equation}
s_{i}^{2}=\frac{-8m_{i}C\left( \sum a_{j}\right) }{a_{i}\hbar ^{2}},
\label{17}
\end{equation}
and 
\begin{equation}
s_{i}^{2}a_{i}m_{j}=s_{j}^{2}a_{j}m_{i}.  \label{18}
\end{equation}
Even if the last formula is valid for any $i$ and $j$, it is trivial unless $%
i\neq j$. This comment applies also to other relationships that like (18)
contain $i$ and $j$ that we will deal with throughout the rest of this
paper. Now, (15) and (18) lead to a consistency condition implying that $%
m_{1}=m_{2}=...=m_{n}=m$. This is somewhat reminiscent of superselection
rules \cite{Cis}. However, whereas the superselection rules stem from some
kinematical or dynamical symmetries, the rule in question is the consequence
of the assumption that particles are uncorrelated and separable and forbids
free of diferent masses to belong to the same quantum-mechanical Universe.
Thus, this rule is much stronger than the superselection rules of linear
quantum mechanics that merely exclude certain types of superpositions, such
as between particles of different mass or between different species of
particles as, for instance, bosons and fermions. The latter exclusion is
called the univalence superselection rule. We suggest that this new type of
rules be called supersuperselection rules.\footnote{%
Or simply $S^{2}S$ rules.} The mass, the univalence, and the particle number
along with other superselection rules are believed to be firmly established
in linear quantum mechanics although objections against some of them have
been raised in the literature \cite{Mir}.

Moreover, (17) entails that all $a_{i}$'s have to be of the same sign or at
least one of $s_{i}^{2}$'s will be negative, and that $C=-|C|$. Combining
(17) and (15) produces another consistency condition, 
\begin{equation}
\sum_{i=1}^{n}a_{i}=\pm \frac{\hbar ^{2}}{8m|C|},  \label{19}
\end{equation}
with the sign of the RHS\ of (19) depending on the sign of $a_{i}$'s, but
otherwise $a_{i}$'s being unspecified. It is also a straightforward
implication of (17) that 
\begin{equation}
\sum_{i=1}^{n}\frac{1}{s_{i}^{2}}=\frac{\hbar ^{2}}{8m|C|}  \label{20}
\end{equation}
which has important consequences for the shape of the probability density of
a three-dimensional particle, as we will see later on.

The energy equation (2) together with (15-18) gives 
\begin{equation}
\hbar \frac{\partial c(t)}{\partial t}=-\frac{1}{2}\sum_{i=1}^{n}mv_{i}^{2}-%
\frac{\hbar ^{2}\left( \sum a_{j}\right) }{2ms_{i}^{2}a_{i}}=-\frac{1}{2}%
m\sum_{i=1}^{n}v_{i}^{2}-\frac{1}{n}\sum_{i=1}^{n}\frac{\hbar ^{2}\left(
\sum a_{j}\right) }{2ms_{i}^{2}a_{i}}.  \label{21}
\end{equation}
Now, as seen from (2-3), the energy $E=-\hbar \left\langle \partial
S/\partial t\right\rangle $, where $<>$ denotes the expectation value of the
quantity embraced. Therefore, we find that for $R_{n}$ normalized to unity, 
\begin{equation}
E=\frac{1}{2}\sum_{i=1}^{n}mv_{i}^{2}+\frac{1}{n}\sum_{i=1}^{n}\frac{\hbar
^{2}\left( \sum a_{j}\right) }{2ms_{i}^{2}a_{i}}=E_{kin}+\frac{\hbar ^{4}}{%
16m^{2}|C|}.  \label{22}
\end{equation}

What we have arrived at is quite remarkable: the system consisting of $n$
objects, in principle different, even if of the same mass, seemingly weakly
non-separable due to the $\left( \Delta S\right) ^{2}$-coupling between them
turns out to be separable and the energy of the system is just the sum of
the kinetic energy of its constituents and some constant self-energy term.
In the subrelativistic approach, this constant term is identified with the
rest mass-energy of the system of $n$ particles equal $nmc^{2}$. In a more
general nonrelativistic scheme, the term in question represents the internal
energy of this system not necessarily equal to its rest mass-energy. In
either approach though, this term is, surprisingly enough, independent of
the number of particles; it is some constant, the same for one particle and
for a gazillion of them.

For identical particles the above formulas become somewhat simpler. For
instance, (17) turns into 
\begin{equation}
s^{2}=\frac{8nm|C|}{\hbar ^{2}},  \label{23}
\end{equation}
and since $M=nm$ is the total mass of the Universe filled with the identical
solitonic particles, we see that the size of the particle, a local quantity,
is determined by the mass of the entire Universe! One can view it as a truly
Machian effect \cite{Bar}. To recall, Mach ca. 1883 conjectured that the
inertia of an object, a local property, is due to the rest of the Universe,
the foremost global object. This came to be known as Mach's principle, and
the situations where a local physical quantity is related to some global one
can be thought of as manifestations of the generalized Mach principle. As we
see, in the case under discussion, the physical size of one particle defined
as the width of its Gaussian probability density, 
\begin{equation}
L_{ph}=\sqrt{2}s=\frac{4\sqrt{nm|C|}}{\hbar }=\frac{4\sqrt{M|C|}}{\hbar },
\label{24}
\end{equation}
is affected by the presence of other particles even though they remain
uncorrelated, and its square is directly proportional to the mass of the
Universe. In reality, a free single particle is never alone, it is part of a
larger entity, the Universe. Assuming that the Universe consists of $n$ such
non-interacting particles, the formalism of linear quantum mechanics allows
us to analyze the equation for the single particle independently of others
since the Schr\"{o}dinger equation for the entire Universe can be separated
into $n$ equations for each of its constituents. There are no side effects
to this formal procedure. A similar procedure, as just demonstrated, can be
applied to the case of $n$ particles in the modification under study. Yet,
in this case the presence of other particles affects the physical size of
each separate particle. In the limit of large $n$, their size would become
macroscopic, even gigantic. The experiment however suggests that quantum
particles do not come in sizes greater than some $l_{\max }$ that is assumed
to be microscopic. To meet this experimental constraint, $C$ should not be
greater than $l_{\max }^{2}\hbar ^{2}/M$, where $M$ stands for the total
mass of identical particles in a model Universe consisting of only one kind
of particles. As already observed, free particles of different masses
constitute separate Universes in our scheme. Consequently, if one assumes
that there is a soliton associated with every free quantum system in this
Universe and that their number is large, but their size is small as it is in
the actual Universe we live in where microscopic quantum particles abound,
then this constant becomes unimaginably small and the theory becomes
``almost linear.'' Nevertheless, the solitons persist. On the other hand, if
we could somehow determine (the upper bound on) the coupling constant $C$,
then knowing the size of one particle (say $s$) we would be able to tell
(the lower bound on) the mass of the model Universe of our modification!

Let us now discuss the case of $n$ one-dimensional particles coupled to
harmonic oscillators. The wave function for this system of particles is
taken in the form of (13-14) and the potential is assumed to be 
\begin{equation}
V(x_{1},x_{2},...,x_{n},t)=\sum_{i=1}^{n}k_{i}(x_{i}-v_{i}t)^{2}=\frac{1}{2}%
\sum_{i=1}^{n}m_{i}\omega _{i}^{2}(x_{i}-v_{i}t)^{2}.  \label{25}
\end{equation}
The energy equation (2) is separable if and only if 
\begin{equation}
b_{i}=m_{i}/\hbar  \label{26}
\end{equation}
and 
\begin{equation}
s_{i}^{4}=\frac{2\hbar ^{2}}{2a_{i}^{2}\hbar ^{2}+\frac{1}{2}m_{i}^{2}\omega
_{i}^{2}}.  \label{27}
\end{equation}
The continuity equation (1) is soluble for each $x_{i}$ separately provided 
\begin{equation}
s_{i}^{2}=\frac{-8m_{i}C\left( \sum a_{j}\right) }{a_{i}\hbar ^{2}}
\label{28}
\end{equation}
and 
\begin{equation}
s_{i}^{2}a_{i}m_{j}=s_{j}^{2}a_{j}m_{i}  \label{29}
\end{equation}
for any $i$ and $j$. Equation (28) implies that either all $a_{j}$ are
positive or negative whereas equations (27) and (29) lead to the first
consistency condition, 
\begin{equation}
\left( \frac{s_{i}}{s_{j}}\right) ^{4}=\frac{2\hbar ^{2}a_{j}^{2}+\frac{1}{2}%
m_{j}^{2}\omega _{j}^{2}}{2\hbar ^{2}a_{i}^{2}+\frac{1}{2}m_{i}^{2}\omega
_{i}^{2}}=\left( \frac{m_{i}}{m_{j}}\frac{a_{j}}{a_{i}}\right) ^{2},
\label{30}
\end{equation}
valid for any $i$ and $j$ which in the free particle case produces the
condition that the masses of particles must be equal. Now, however, this
condition does not have to be met, which, in other words, means that the
potential can remove the degeneration in the mass. This last formula can be
transformed into 
\begin{equation}
\left[ \left( \frac{m_{i}}{m_{j}}\right) ^{2}-1\right] =\frac{%
m_{j}^{2}\omega _{j}^{2}}{4\hbar ^{2}a_{j}^{2}}\left[ 1-\left( \frac{m_{i}}{%
m_{j}}\right) ^{4}\left( \frac{a_{j}}{a_{i}}\right) ^{2}\left( \frac{\omega
_{i}^{2}}{\omega _{j}^{2}}\right) ^{2}\right] .  \label{31}
\end{equation}
This constitutes the equation from which to determine $a_{j}^{2}$ as a
function of the other parameters. In what follows, we will consider only the
simplest case that corresponds to $m_{i}=m_{j}=m$ for any $i$ and $j$. Then 
\begin{equation}
\frac{a_{j}}{a_{i}}=\pm \frac{\omega _{j}}{\omega _{i}}  \label{32}
\end{equation}
for arbitrary $i$ and $j$. Consequently, 
\begin{equation}
s_{i}^{2}=-\frac{8Cm\sum \omega _{j}}{\hbar ^{2}\omega _{i}},  \label{33}
\end{equation}
where $C=-|C|$. The last formula and (27) lead to another consistency
condition 
\begin{equation}
a_{1}^{2}=\omega _{1}^{2}\left[ \frac{\hbar ^{4}}{64C^{2}m^{2}\left( \sum
\omega _{j}\right) ^{2}}-\left( \frac{m}{2\hbar }\right) ^{2}\right]
\label{34}
\end{equation}
which essentially is an equation for $a_{1}$ in terms of the parameters of
the system. Knowing the value of $a_{1}$ one can determine other $a_{i}$'s
from (32). Moreover, since $a_{1}^{2}>0$ (the case of $a_{1}^{2}=0$
corresponds to linear quantum mechanics), we obtain that 
\begin{equation}
\sum_{i=1}^{n}\omega _{i}<\frac{\hbar ^{3}}{4|C|m^{2}}.  \label{35}
\end{equation}
This formula replaces (19) valid for free particles, but (20) still holds in
this case. Now, the energy equation (2) yields 
\begin{equation}
\hbar \frac{\partial c(t)}{\partial t}=-\frac{1}{2}\sum_{i=1}^{n}mv_{i}^{2}-%
\frac{\hbar ^{2}\left( \sum a_{j}\right) }{2ms_{i}^{2}a_{i}}-\frac{1}{2}%
\sum_{i=1}^{n}m\omega _{i}^{2}s_{i}^{2}  \label{36}
\end{equation}
so that the total energy of the system is 
\begin{equation}
E=-\hbar \left\langle \frac{\partial S}{\partial t}\right\rangle =E_{kin}+%
\frac{\hbar ^{4}}{16m^{2}|C|}+\frac{4|C|m^{2}}{\hbar ^{2}}\left(
\sum_{i=1}^{n}\omega _{i}\right) ^{2}<E_{kin}+\frac{5\hbar ^{4}}{16m^{2}|C|}.
\label{37}
\end{equation}

We observe that the wave function of one particle is affected not only by
the frequency of the oscillator it is coupled to but also by the frequencies
of other oscillators. One can thus wonder if this violates causality. One
can imagine that, for simplicity, in a system of two particles coupled to
oscillators of different frequencies changing the frequency of one of them
will affect the probability density of the particle coupled to the other
one. The change in question means a global change in the potential of one of
the particles as the potential is affected in its entire domain. In this
sense, this differs from a telegraph that employs spins on which operations
are local. Nevertheless, in a system like that one can transmit information
instantaneously from one observer to another one separated by an arbitrary
distance and thus the causality in this system can indeed be violated. The
energy changes in this process, but one cannot tell whether it is
transmitted or not for it is not localized.

This is not necessarily unexpected for, as already pointed out in \cite
{Pusz1}, the equations of motion are not separable in the weak sense \cite
{Bial} that assumes a factorizable wave function for a compound system. On
the other hand, the discussed model constitutes an elementary example of a
causality violation mechanism in nonlinear quantum mechanics which does not
involve the spin \cite{Gis, Pol, Czach1, Czach2} or any complex reasoning.
If only because of that, it may deserve some attention. It should also be
noted that situations like that are rather generic in nonlinear systems.
Nevertheless, the consensus is that situations of this kind are unphysical
and because of that they are dismissed.

The problem in question can, in fact, be alleviated within the theory itself
by selecting the parameters of the solution in such a way that the
demonstrated causality violation does not occur. One does that by assuming
that all $a_{i}$'s are equal. As a result, (30) implies now that 
\begin{equation}
\omega _{j}^{2}=\left( \frac{m_{1}}{m_{j}}\right) ^{4}\omega _{1}^{2}+2\hbar
^{2}a^{2}\left[ \left( \frac{m_{1}}{m_{j}}\right) ^{2}-1\right] ,  \label{38}
\end{equation}
whereas (27) and (28) yield that 
\begin{equation}
\omega _{i}^{2}=\frac{4\hbar ^{6}}{64m_{i}^{4}C^{2}n^{2}}-\frac{4a^{2}\hbar
^{2}}{m_{i}^{2}}.  \label{39}
\end{equation}
Applying the last formula for $i=1$ and using it in (38) gives 
\begin{equation}
\omega _{j}^{2}=\frac{4\hbar ^{6}}{64m_{j}^{4}C^{2}n^{2}}-\frac{%
4m_{1}^{2}a^{2}\hbar ^{2}}{m_{j}^{4}}+2\hbar ^{2}a^{2}\left[ \left( \frac{%
m_{1}}{m_{j}}\right) ^{2}-1\right]  \label{40}
\end{equation}
which is equal to (39) specialized for $i=j$ only if $m_{j}=m_{1}$. In view
of the generality of this argument one is lead to the conclusion that all
the masses $m_{i}$ must be the same and so the particles-oscillators are
identical. The parameter $a$ is determined by the frequency $\omega $ and
other constants, 
\begin{equation}
a^{2}=\frac{\hbar ^{4}}{64m^{2}C^{2}n^{2}}-\frac{m^{2}\omega ^{2}}{4\hbar
^{2}}.  \label{41}
\end{equation}
As seen from the last formula or (35), the frequency, which is the same for
all oscillators, has to satisfy the condition 
\begin{equation}
\omega <\frac{\hbar ^{3}}{4|C|nm^{2}}  \label{42}
\end{equation}
corresponding to $a^{2}>0$. The energy of the field of oscillators is now 
\begin{equation}
E=E_{kin}+\frac{\hbar ^{4}}{16m^{2}|C|}+\frac{4|C|m^{2}n^{2}}{\hbar ^{2}}%
\omega ^{2}<E_{kin}+\frac{5\hbar ^{4}}{16m^{2}|C|}.  \label{43}
\end{equation}
The picture that emerges from this is that of a field of $n$ identical
particles oscillating in unison. Changing the frequency of a single
oscillator is possible only if the same change takes place in the entire
field. This can happen instantaneously but does not lead to any
quantum-mechanical violation of causality because the probability density of
the particles is not changed in the process, as opposed to their phase. The
situation in question is somewhat similar to the EPR phenomenon \cite{Ein,
Bohm} in that the observer measuring or changing the frequency of one
oscillator knows what the frequency of other oscillators is going to be at
the very same moment. However, unlike in the standard EPR experiment, the
nonlocal impact of measurement is not confined to just one particle but has
a global sweep. Moreover, the particles are not entangled as is the case for
the EPR type of correlations. Yet, by changing the frequency of her
oscillator, one observer can change the frequency of all the other
oscillators in the Universe. If there are any observers associated with
them, they will notice this change. It is obvious that one can
instantaneously transmit information in this way, although it is not clear
if the same is true about the energy for the latter is not localized. It
does change though in this process.

It should be stressed that the discussed solitonic solution does not
invalidate nor replaces the multi-oscillator solutions of linear quantum
mechanics. Such solutions still exist in the modification under study.
However, if the wave function of the Universe is to be factorizable, the
oscillators of linear theory cannot share this Universe with their solitonic
brethren. A wave function that would accomodate both kinds of oscillators is
not supported by the equations of motion as seen from (29). This relation
cannot be satisfied if one of $a_{i}$'s is zero which would correspond to a
``linear'' quantum oscillator. Either all $a_{i}$'s must be zero or none. As
a matter of fact, even the free solitons and the solitons coupled to
harmonic oscillators cannot belong to the same Universe if they are to be
uncorrelated. Let us now demonstrate this fact. Without any loss of
generality, we can assume that the Universe contains only one free particle,
all others being oscillators. Putting $\omega _{1}=\omega _{i}=0$ in (30)
and solving it for $a_{j}$ gives 
\begin{equation}
a_{j}=\pm \frac{m_{j}\omega _{j}}{2\hbar \sqrt{1-\eta ^{2}}},  \label{44}
\end{equation}
where we need to assume that $\eta =m_{1}/m_{j}<1$. $\eta $ equal $1$
implies that $\omega _{j}=0$ and so is not interesting, but this observation
will prove useful later on. Inserting this into (27) leads to 
\begin{equation}
s_{j}^{4}=\frac{4\hbar ^{2}\left[ 1-\eta ^{2}\right] }{m_{j}^{2}\omega
_{j}^{2}\left[ 2-\eta ^{2}\right] }.  \label{45}
\end{equation}
Now, from (27) and (28) applied to the free particle we obtain that 
\begin{equation}
1=\left( s_{1}^{2}a_{1}\right) ^{2}=\left( \frac{8m_{1}|C|}{\hbar ^{2}}%
\right) ^{2}\left( \sum a_{j}\right) ^{2}.  \label{46}
\end{equation}
Applying (28) to the $j$-th oscillator and using (46) and (44) yields 
\begin{equation}
s_{j}^{4}=\left( \frac{8m_{j}|C|}{a_{j}\hbar ^{2}}\right) ^{2}\left( \sum
a_{j}\right) ^{2}=\frac{m_{j}^{2}}{m_{1}^{2}a_{j}^{2}}=\frac{4\hbar
^{2}\left[ 1-\eta ^{2}\right] }{m_{1}^{2}\omega _{j}^{2}}.  \label{47}
\end{equation}
If (45) and (47) are to be equal, $\eta $ must be$1$, which means that the $%
j $-th particle is also free. Therefore, we conclude that the factorizable
wave function of the Universe can describe either only the free or only the
coupled solitons, but not both those species together.

Finally, following the scheme for the $n$ one-dimensional particles, it is
straightforward to find a $d$-dimensional one-particle solitonic solution.
The wave function of such a particle is given by (13-14) with $n$ replaced
by $d$. The energy of a single $d$-dimensional quanton is 
\begin{equation}
E=\frac{1}{2}\sum_{i=1}^{d}mv_{i}^{2}+\frac{\hbar ^{2}}{2s^{2}}d=E_{kin}+%
\frac{\hbar ^{4}}{16m^{2}|C|},  \label{48}
\end{equation}
where 
\begin{equation}
s^{2}=\frac{8d|C|m}{\hbar ^{2}}.  \label{49}
\end{equation}
The last formula assumes that the particle is spherical. One can now treat
even a more general case of $n$ particles in $d$ dimensions, which would
remove the assumption of one-dimensional particles. Instead of (49) we can
also use (20) to obtain a greater variety of shapes for one
three-dimensional particle. For instance, if, say, $s_{1}$ and $s_{2}$ are
much greater than $s_{3}$ we obtain a sheet-like configuration extended in
the $x_{1}$-$x_{2}$ plane, while for $s_{1}$ and $s_{2}$ being small but of
the same order and $s_{3}$ being much greater we obtain a string-like
configuration extended along the direction of $x_{3}$. Even if these
configurations have different shapes they are topologically equivalent to a
three-dimensional spherical particle. Moreover, they all have the same
energy.

Finally, let us note that if the multi-particle equations are derived from
the Lagrangian \cite{Pusz1} 
\begin{equation}
-L_{n}(R,S)=\hbar R^{2}\frac{\partial S}{\partial t}+\sum_{i=1}^{n}\frac{%
\hbar ^{2}}{2m_{i}}\left[ \left( \vec{\nabla}_{i}R\right) ^{2}+R^{2}\left( 
\vec{\nabla}_{i}S\right) ^{2}\right] +CR^{2}\sum_{i=1}^{n}\left( \Delta
_{i}S\right) ^{2}+R^{2}V,  \label{50}
\end{equation}
they are weakly separable and do not lead to any strange nonlocal effects in
the systems we examined above. In particular, one can assume here that $%
C=C_{i}$, i.e., the coupling constant is particle-dependent, similarly as
the mass. For $n=1$, the Lagrangian in question reduces to the same form as
the Lagrangian of (4), but in the multi-particle case these Lagrangians
describe different theories. The nonlocal effects may however appear when
nonfactorizable wave functions are considered.

\section{Conclusions}

We have presented the $n$-particle free solitonic solution and the solution
describing $n$ particles coupled to harmonic oscillators in the nonlinear
modification of the Schr\"{o}dinger equation put forward recently \cite
{Pusz1}. In addition, a particular solution for two free solitons of the
same mass that does not stem from this general scheme has also been
presented. This solution represents two particles whose compound wave
function does not factorize and whose kinetic energy is, surprisingly
enough, the same as the kinetic energy of one particle.

The most remarkable property of the free $n$-particle solution is the
remnant inseparability that manifests itself in the wave function of
individual particles depending on the properties of other particles of the
system. In the simplest case of identical particles, it is the total number
of particles that appears in the wave function of an individual particle. As
a result, the width of each particle's Gaussian probability density, a local
property, is directly linked to the mass of the entire Universe in a very
Machian manner. Nevertheless, they all evolve independently. In the case of
solitons coupled to harmonic oscillators the probability density of one
particle is affected not only by the frequency of its own oscillator, but by
the frequencies of other oscillators as well. This leads to a causality
violation as any change in the frequency of one of the oscillators is
instantaneously detected by the observers related to other oscillators. By
measuring the probability density of the particles coupled to their
oscillators, they realize that someone is ``messing'' with these particles.
We have shown, however, that it is possible to avoid this particular problem
by an appropriate choice of the parameters of the solution. Namely, if the
ratio of $a_{i}/a_{j}$ is one, the nonseparability becomes somewhat similar
to the EPR kind, but without the entanglement typical for the EPR
correlations. The probability density of a single oscillator is then not
related to what is happening to the other oscillators, but they all have to
oscillate with the same frequency and so in this sense they remain
correlated. It is possible to transmit information in this way, just by
changing the frequency of one oscillator, similarly as in the more generic
case. The simplest way to avoid the discussed nonlocality problems is to
adopt a weakly separable multi-particle extension of the basic one-particle
nonlinear equation, although this does not guarantee that similar problems
will not occur for nonfactorizable wave functions.

It would be interesting to see how one can circumvent this problem in the
strong separability framework advocated by Czachor \cite{Czach3} that
originated in the work of Polchinski \cite{Pol} and Jordan \cite{Jord}. This
approach chooses as its starting point the nonlinear von Neumann equation
for the density matrices \cite{Czach4} and proceeds from there to the $n$%
-particle extension. In a sense, one can call this approach ``effective'' as
opposed to the ``fundamentalist'' one. The latter approach does not resort
to reformulating in terms of density matrices various proposed in the
literature nonlinear modifications of the Schr\"{o}dinger equation that,
similarly as the linear equation itself, are postulated to be obeyed by pure
states. The problem of nonlocality is confounded since these approaches,
even though they belong to the same strong separability framework, yield
different results. For instance, the Bia\l ynicki-Birula and Mycielski
modification that is both weakly and strongly separable in the effective
approach turns out to be essentially nonlocal in the fundamentalist approach
for a general nonfactorizable two-particle wave function, as recently
demonstrated by L\"{u}cke \cite{Luc1}. The same applies to the
Doebner-Goldin modification that being weakly separable and strongly in
the effective approach fails to maintain the separability for a more general
nonfactorizable wave function describing a system of two particles if the
fundamentalist approach is used. Only in some special case, that corresponds
to the linearizable Doebner-Goldin equation, are the particles separable in
the latter approach \cite{Luc2}.

One of the consequences of the discussed residual inseparability of free
particles is the smallness of the coupling constant $C$. In the simplest
case of identical particles, the width of the probability density of a single
particle is proportional to the total number of particles and the constant
in question. In the limit of large $n$ which is the most natural to consider
if one assumes that every quantum system is associated with a particle, this
width can be macroscopic if not just gigantic unless $C$ is really very
small. Since the experiments do not support the view that quantum particles
are of macroscopic size, the nonlinear coupling constant must indeed be very
small. One needs to realize though that this is only a model which rests on
the assumptions that there exist a particle associated with every quantum
system, that quantum systems are microscopic, and that all of them have the
same mass. None of these assumptions have to be true, but if there existed a
Universe where these assumptions were correct then the coupling constant of
the discussed nonlinear extension of the Schr\"{o}dinger equation would be
incredibly small.

A novel property that also emerged from the free multi-particle solution is
an exclusion rule that forbids particles of different masses to be part of
the same Universe if they are to be uncorrelated and separated. The same
holds true for the solitonic oscillators of common $a$ which are supposed to
share the frequency. This is somewhat characteristic of superselection rules
that stem from the symmetry considerations, but is much stronger in its
implications for the standard superselection rule applicable in this case
only forbids superpositions of quantum mechanical systems of different
masses. We propose that the rules of this type be called supersuperselection
rules. Similar to it is the univalence supersuperselection rule that under
the same conditions prevents different species of particles, such as the
free solitons and the solitonic oscillators in our case, from belonging to
the same Universe. The same rule applies also to the oscillators of linear
quantum mechanics and the ``nonlinear'' oscillators of the modified
Schr\"{o}dinger equation. Similarly as the nonlocal effects, the rules in
question are native to only one of the discussed $n$-particle extensions;
they do not emerge if the multi-particle equations are weakly separable.

\section*{Acknowledgments}

I would like to thank Professor P. O. Mazur for bringing my attention to the
work of Professor Staruszkiewicz that started my interest in nonlinear
modifications of the Schr\"{o}dinger equation. A correspondence with Dr. M.
Czachor and his comments on the preliminary versions of this paper are
gratefully acknowledged as is a correspondence with Professor Wolfgang
L\"{u}cke concerning his recent work. This work was partially supported by
the NSF grant No. 13020 F167 and the ONR grant R\&T No. 3124141.

\end{document}